\documentclass[conference,romanappendices]{IEEEtran}
\usepackage{CCE}
\IEEEoverridecommandlockouts
\usepackage[left=1.62cm,right=1.62cm,top=1.9cm]{geometry}
\newcommand{\CLASSINPUTtoptextmargin}{2.4cm}
\newcommand{\CLASSINPUTbottomtextmargin}{4.6cm}
\usepackage{silence} 
\usepackage[T1]{fontenc}
\usepackage[latin9,utf8]{inputenc}
\usepackage{amsmath,amsfonts,amsthm,amssymb}
\usepackage{algorithmic,algorithm}
\usepackage{array,stackrel}
\usepackage[caption=false,font=normalsize,labelfont=sf,textfont=sf]{subfig}
\usepackage{textcomp}
\usepackage{stfloats}
\usepackage{url}
\usepackage{verbatim}
\usepackage{graphicx}
\usepackage{cite}
\usepackage{xcolor}
\usepackage{pdfcolmk}
\usepackage{refcount}
\usepackage{esint}
\usepackage{hhline}
\usepackage{bm}
\usepackage{xfrac}
\usepackage{makecell}
\usepackage{float}
\usepackage{lipsum}
\usepackage{hyperref}
\usepackage{cases}
\usepackage[]{footmisc}
\def\BibTeX{{\rm B\kern-.05em{\sc i\kern-.025em b}\kern-.08em
    T\kern-.1667em\lower.7ex\hbox{E}\kern-.125emX}}

\DeclareRobustCommand{\foreign@language}[1]{%
  \lowercase{\oldforeign@language{#1}}}
\theoremstyle{plain}
\newtheorem{thm}{\protect\theoremname}
\theoremstyle{plain}
\newtheorem{lem}[thm]{\protect\lemmaname}
\newtheorem{corollary}{Corollary}[thm]
\usepackage{stackengine}

\usepackage[caption=false,font=normalsize,labelfont=sf,textfont=sf]{subfig}

\makeatother

\providecommand{\lemmaname}{Lemma}
\providecommand{\theoremname}{Theorem}

\usepackage{tabularx,booktabs}
\newcolumntype{C}{>{\centering\arraybackslash}X} % centered version of "X" type
\definecolor{DR}{rgb}{0.7,0,0}
\definecolor{DG}{rgb}{0,0.5,0}
\definecolor{DB}{rgb}{0,0,0.7}
\begin{document}

\title{Outage performance of the $\alpha$-Beaulieu-Xie Shadowed Fading Channel Model
% \thanks{This work was supported by the Russian Science Foundation under grant 22-29-01458 (https://rscf.ru/en/project/22-29-01458/)}
%\thanks{978-1-6654-0306-1/21/\$31.00 ©2023 IEEE}
}

 \author{\IEEEauthorblockN{Aleksey~S.~Gvozdarev,~\IEEEmembership{Member,~IEEE}}
 \IEEEauthorblockA{\textit{Department of Intelligent Radiophysical Information Systems (IRIS)} \\
 \textit{P.G.~Demidov Yaroslavl State University}\\
 Yaroslavl, Russia \\
 asg.rus@gmail.com, {\includegraphics[scale=0.1]{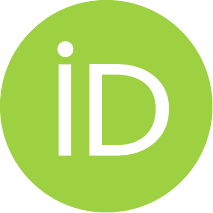}}~0000-0001-9308-4386}
\thanks{\copyright 2023 IEEE.  Personal use of this material is permitted.  Permission from IEEE must be obtained for all other uses, in any current or future media, including reprinting/republishing this material for advertising or promotional purposes, creating new collective works, for resale or redistribution to servers or lists, or reuse of any copyrighted component of this work in other works.}
\thanks{This work was supported by Russian Science Foundation under Grant 22-29-01458 (https://rscf.ru/en/project/22-29-01458/).}%
}

%\author{Aleksey S.~Gvozdarev\href{https://orcid.org/0000-0001-9308-4386}{\includegraphics[scale=0.1]{orcid.pdf}},~\IEEEmembership{Member,~IEEE,}
%        % <-this % stops a space
%\thanks{\copyright 2023 IEEE.  Personal use of this material is permitted.  Permission from IEEE must be obtained for all other uses, in any current or future media, including reprinting/republishing this material for advertising or promotional purposes, creating new collective works, for resale or redistribution to servers or lists, or reuse of any copyrighted component of this work in other works.}
%\thanks{The author is with the Department of Intelligent Radiophysical  Information Systems, P. G. Demidov Yaroslavl State University, 150003 Yaroslavl,
%Russia (e-mail: asg.rus@gmail.com).}% <-this % stops a space
%\thanks{This work was supported by Russian Science Foundation under Grant 22-29-01458 (https://rscf.ru/en/project/22-29-01458/).}%
%}

\maketitle

\begin{abstract}
The research presents the closed-form outage analysis of the newly presented $\alpha$-modification of the shadowed Beaulieu-Xie fading model for wireless communications. For the considered channel, the closed-form analytical expressions for the outage probability (including its upper and lower bounds), raw moments, amount of fading, and channel quality estimation indicator are derived. The carried out thorough numerical simulation and analysis demonstrates strong agreement with the presented closed-form solutions and illustrates the relationship between the outage probability and channel parameters.
\end{abstract}


% \fi
\IEEEpeerreviewmaketitle

\section{Introduction}\label{S-1}
\IEEEPARstart{T}{he} progress in the ad hoc communications design  nowadays is mostly limited by the existing restrictions implied by the wireless propagation channels used \cite{Hen23, Sil23}, especially in the millimeter-wave range, terahertz (THz) bands (including free-space optical (FSO) communication). This means that the more adequate is the channel model to the real-life propagation conditions, the better is the overall system performance description and prediction.  This means that within the general framework of the communication system's analysis, the primary role is given to the channel model, which imposes specific restrictions on the employed signal processing and system parameters.

To achieve such adequacy, certain model description complication is needed (i.e., to include finer propagation effects), which inherently leads to the increasing complexity of its mathematical description. Thus, in practice, to keep the description analytically tractable, a feasible model with moderate complexity is sought.

As a possible widely used candidate, one can mention the Beaulieu-Xie (BX) model that was proposed in \cite{Bea15}. It assumes a multiple Line-of-Sight (LoS) propagation (being of paramount importance for THz and FSO communication). Up to now, the BX channel has been numerously used to assess the  communication systems' performance, including their ergodic capacity \cite{Kan17}; level crossing rate and average fade duration \cite{Olu17}; joint envelope-phase statistics \cite{Sil21}; energy efficiency for various modulation schemes  \cite{Kan18,Kan20, Bil22}; diversity reception with maximum ration combining \cite{Kau20} and switch-and-stay \cite{Sha22} receivers; the classical physical layer security (PLS) \cite{Cha20} and  \cite{Sun22} PLS with non-orthogonal multiple access with in-phase/quadrature imbalance.

It was further modified by A. Olutayo et al. (see \cite{Olu20a})  by incorporating shadowing effects of the LoS components. The particular trait of the resultant BX-shadowed model is its capability of managing arbitrary positive shadowing coefficients, in contrast to the classical BX-model, where it is lower-bounded by 0.5 (due to the usage of the Nakagami shadowing). The BX-shadowed model was employed in the studies concerning the communication system's bit/symbol error rate \cite{Haw22}, physical layer security \cite{Gvo22a, Gvo22b}, and channel capacity \cite{Sil22a}.

Specifically for high-frequency communications (THz and FSO) \cite{Upa17, Isl20, Li22a}, it is common to derive the so-called $\alpha$-modifications of the existing models, which are assumed to include potential nonlinear propagation effects. Originating from the work by M.D.~Yacoub \cite{Yac07b} introducing the physical model of the Stacy distribution, and up to the most recent ones \cite{Sil22b, Sil22c, AlH22a, AlH22b}, they got a wide acceptance in the modern communication systems \cite{Bha21, Li22}.

Lately, $\alpha$-modifications of the initial  BX \cite{Che23} and the BX-shadowed \cite{Gvo23} models were proposed. They presented the exact expressions for the basic probability functions (i.e., density function (pdf) and distribution function (cdf)), however, no  further statistical description is available. Furthermore, the expressions presented in \cite{Che23} incorporated the initial  restrictions of the classical model (i.e., integer-valued fading parameter's limitation). Moreover, none of the works presented the outage analysis.

The proposed research expands the framework adopted in the  recently published paper \cite{Gvo23} and withdraws those deficiencies, presenting an outage analysis of the $\alpha$-Beaulieu-Xie shadowed fading channel model. The main contributions of the research are as follows: \textit{a)} an  outage analysis  of the $\alpha$-BX-shadowed channel; \textit{b)} the derived expression for the instantaneous signal-to-noise ratio (SNR) raw moments, amount of fading, and channel quality estimation indicator; \textit{c)} the proposed analytical description is illustrated on the problem of outage probability calculation, and its lower and upper bounds are evaluated, and used for the proposed joint quality-reliability analysis; \textit{d)} a comprehensive numerical study verifying the correctness of the derived expressions is  carried out.

The submission is structured as follows: Section~\ref{S-2} delivers a concise existing probabilistic description of the channel model under consideration (initial BX shadowed model and its $\alpha$ modification) as well as the novel results obtained in the research (e.g., raw moments, amount of fading, channel quality estimation index, outage probability with its lower and upper bounds, and joint quality-reliability analysis); Section~\ref{S-3} contains the results (and their discussion)  of a comprehensive numerical outage analysis for all possible channel parameters; with the conclusions in Section~\ref{S-4}.

\section{Channel model}\label{S-2}
To perform the outage analysis of the model under consideration, one needs to introduce the classical Beaulieu-Xie shadowed model, its $\alpha$-modification, and basic statistical description of the latter one.

\subsection{Classical Beaulieu-Xie shadowed model}\label{Ss-2-1}
The initial Beaulieu-Xie shadowed model (proposed in \cite{Olu20a}) is a four-parametric model with the envelope pdf expressed in the following form:
\begin{IEEEeqnarray}{rCl}\label{eq:1}
f_{\bar{R}}(\bar{r})&=&\frac{2 \bar{r}^{2m_X-1}}{\Gamma (m_X)}\left(\frac{m_Y\Omega_X}{m_Y\Omega_X+m_X\Omega_Y}\right)^{m_Y} \left(\frac{m_X}{\Omega_X}\right)^{m_X}   \times    \IEEEnonumber \\
&&\hspace{-20pt}   \times   \mbox{}_1F_1\left(m_Y, m_X, \frac{m_X^2\Omega_Y \bar{r}^2}{\Omega_X(m_Y\Omega_X+m_X\Omega_Y)}\right)e^{-\frac{m_X}{\Omega_X}\bar{r}^2},
\end{IEEEeqnarray}
here $\Gamma (   \cdot )$ is the gamma-function \cite{DLMF}, and $\mbox{}_1F_1(   \cdot )$ is the confluent Kummer's  function \cite{DLMF}. The channel parameters (i.e., $m_X, m_Y$, $\Omega_X, \Omega_Y$) control the total signal fading ($m_X$),  LoS component's shadowing ($m_Y$), NLoS and LoS power ($\Omega_X, \Omega_Y$), respectively.

It must be pointed out that it is not lower-limited by the values of fading/shadowing coefficients, and in special simplified conditions degenerates to  a wide range of classical fading models (see \cite{Olu20a}), for instance, the generalized (e.g., $\eta-\mu$, $\kappa-\mu$ shadowed, and $\kappa-\mu$), as well as the simplified ones (e.g., Nakagami-m, one-sided Gaussian, Rayleigh, Rician shadowed, Rician) etc.

\subsection{$\alpha$-modification of the Beaulieu-Xie shadowed model: prior work}\label{Ss-2-2}

The closed-form derivations of the $\alpha$-modification for the BX-channel model \eqref{eq:1} (performed in \cite{Gvo23}) assumed the concept proposed by M.D.~Yacoub (see \cite{Yac07b}).  Within such an approach, one assumes that the signal propagates in a nonhomogeneous environment  and undergoes nonlinear envelope impairments.

The expressions for the basic first-order statistical description are derived without any restrictions implied on the channel parameters $m_X, m_Y, \Omega_X, \Omega_Y, \alpha$, and are given by the following Lemma.
\begin{lem} The basic probabilistic description of the $\alpha$-BX-shadowed channel with arbitrary  parameters ($m_X, m_Y, \Omega_X, \Omega_Y, \alpha \geq 0$), including pdf $f_{\gamma }(\gamma )$ and  cdf $F_{\gamma }(\gamma )$  of the instantaneous SNR $\gamma$ are given in the following form:
\begin{IEEEeqnarray}{rCl}\label{eq:2}
f_\gamma (\gamma )&=&\frac{\alpha \left(\frac{\gamma }{\bar{\gamma }}\right)^{\frac{\alpha m_X}{2}-1} \left(\frac{m_Y\Omega_X}{m_Y\Omega_X+m_X\Omega_Y}\right)^{m_Y}}{2\mathrm{C}_{\alpha }^{m_X}\Gamma (m_X)\bar{\gamma }e^{\frac{1}{\mathrm{C}_{\alpha }}\left(\frac{\gamma }{\bar{\gamma }}\right)^{\frac{\alpha }{2}}}}    \times    \IEEEnonumber \\
&&\hspace{-40pt}   \times   \mbox{}_1F_1\left(m_Y, m_X, \frac{1}{\mathrm{C}_{\alpha }}\left(\frac{m_X\Omega_Y}{m_Y\Omega_X+m_X\Omega_Y}\right)\left(\frac{\gamma }{\bar{\gamma }}\right)^{\frac{\alpha }{2}}\right),
\end{IEEEeqnarray}
\begin{IEEEeqnarray}{rCl}\label{eq:3}
F_\gamma (\gamma )&=&\frac{\left(\frac{m_Y\Omega_X}{m_Y\Omega_X+m_X\Omega_Y}\right)^{m_Y}}{\mathrm{C}_{\alpha }^{m_X}\Gamma (m_X+1)}\left(\frac{\gamma }{\bar{\gamma }}\right)^{\frac{\alpha m_X}{2}} e^{-\frac{1}{\mathrm{C}_{\alpha }}\left(\frac{\gamma }{\bar{\gamma }}\right)^{\frac{\alpha }{2}}}    \times    \IEEEnonumber \\
&&\hspace{-50pt}   \times    \Phi_2\!\!\left(\!1, m_Y; m_X; \frac{1}{\mathrm{C}_{\alpha }}\!\left(\frac{\gamma }{\bar{\gamma }}\right)^{\!\!\frac{\alpha }{2}}\!\!\!, \!\left(\frac{\mathrm{C}_{\alpha }^{-1}m_X\Omega_Y}{m_Y\Omega_X+m_X\Omega_Y}\right)\left(\frac{\gamma }{\bar{\gamma }}\right)^{\!\!\frac{\alpha }{2}}\!\! \right)\!\!,
\end{IEEEeqnarray}
where $\bar{\gamma }$ is the expected value of the instantaneous SNR, $\Phi_2(   \cdot )$ is the bivariate confluent Appell function \cite{DLMF}, and $\mathrm{C}_{\alpha }$ is defined as
\begin{equation}\label{eq:4}
\hspace{-10pt}\mathrm{C}_{\alpha }=\left[\frac{\Gamma (m_X)}{\Gamma \left(m_X+\frac{2}{\alpha }\right)\mbox{}_2F_1\left(m_Y,-\frac{2}{\alpha }; m_X; -\frac{m_X\Omega_Y}{m_Y\Omega_X}\right)}\right]^{\frac{\alpha }{2}},
\end{equation}
with $\mbox{}_2F_1(   \cdot )$ being the Gauss hypergeometric function \cite{DLMF}.
\end{lem}

\begin{IEEEproof}
For the detailed proof, see \cite{Gvo23}.
\end{IEEEproof}

\subsection{$\alpha$-Beaulieu-Xie shadowed model: new derived results}\label{Ss-2-3}
\subsubsection{SNR raw moments}\label{Sss-2-3-1}
The obtained result helps to derive the expressions for the arbitrary-order moments of the SNR.

\begin{thm} The moments $\gamma^k$ of the instantaneous SNR $\gamma$ for the $\alpha$-BX-shadowed model \eqref{eq:2} are given by
\begin{IEEEeqnarray}{rCl}\label{eq:5}
\mathbb{E}\{\gamma^k\}&=&\bar{\gamma }^k\left(\frac{\Gamma (m_X)}{\Gamma \left(m_X+\frac{2}{\alpha }\right)}\right)^k
    \frac{\Gamma \left(m_X+\frac{2k}{\alpha }\right)}{\Gamma \left(m_X\right)}
       \times    \IEEEnonumber \\
  &&   \times    \frac{\mbox{}_2F_1\left(m_Y,-\frac{2k}{\alpha }; m_X; -\frac{m_X\Omega_Y}{m_Y\Omega_X}\right)}{\left(\mbox{}_2F_1\left(m_Y,-\frac{2}{\alpha }; m_X; -\frac{m_X\Omega_Y}{m_Y\Omega_X}\right)\right)^k}.
\end{IEEEeqnarray}
\end{thm}
\begin{IEEEproof}
Combining the definition of the $k$-th order moment (i.e., $\mathbb{E}\{\gamma^k\}$) with the obtained expression \eqref{eq:2} for the pdf in conjunction with the following integration property of the generalized hypergeometric function $\mbox{}_mF_n(   \cdot )$ (see \cite{Gra96} equation $7.522.9$):
\begin{IEEEeqnarray}{rCl}\label{eq:6}
&&\hspace{-5pt}\int_0^{\infty}x^{\sigma-1}e^{-\mu x}\mbox{}_mF_n(a_1, \ldots, a_m; b_1, \ldots, b_m;\lambda x){\rm{d}}x= \IEEEnonumber \\
&&\hspace{10pt}=\Gamma (c)\mu^{-\sigma } \mbox{}_{m+1}F_n\left(a_1, \ldots, a_m, \sigma; b_1, \ldots, b_m;\frac{\lambda }{\mu }\right)
\end{IEEEeqnarray}
after simplifications yields the required result.
\end{IEEEproof}

\subsubsection{Amount of fading and channel quality estimation index}\label{Sss-2-3-2}

The derived expression for the arbitrary-order raw moments is of primary importance for the evaluations of the channel quality estimation index (CQEI) \cite{Bad20}. Being initially proposed in \cite{Lio07} as an alternative long-term performance criterion helping to evaluate and optimize wireless systems functioning in the case of fading, it found wide acceptance in a variety of applications \cite{Alm21, Teg22}. The classical definition of CQEI is given either in terms of the raw moments of the instantaneous SNR or in terms of the amount of fading (AoF), which in its turn is evaluated as the variance of $\gamma$ normalized by $\bar{\gamma }^2$, i.e,

\begin{IEEEeqnarray}{rCl}\label{eq:x1}
&&\mathrm{CQEI} = \frac{\mathbb{V}\mathrm{ar}\{\gamma \}}{\mathbb{E}\{\gamma \}^3}=\frac{\mathrm{AoF}}{\mathbb{E}\{\gamma \}}.
\end{IEEEeqnarray}

Thus, Theorem 2 helps to state the following closed-from result.

\begin{thm} AoF and CQEI for the $\alpha$-BX-shadowed model \eqref{eq:2} are evaluated as:
\begin{IEEEeqnarray}{rCl}
\mathrm{AoF}&=&
    \frac{\Gamma \left(m_X\right)\Gamma \left(m_X+\frac{4}{\alpha }\right)}{\left(\Gamma \left(m_X+\frac{2}{\alpha }\right)\right)^2}
       \times    \IEEEnonumber \\
  &&   \times    \frac{\mbox{}_2F_1\left(m_Y,-\frac{4}{\alpha }; m_X; -\frac{m_X\Omega_Y}{m_Y\Omega_X}\right)}{\left(\mbox{}_2F_1\left(m_Y,-\frac{2}{\alpha }; m_X; -\frac{m_X\Omega_Y}{m_Y\Omega_X}\right)\right)^2}, \label{eq:x2}\\
\mathrm{CQEI}&=&\frac{1}{\bar{\gamma }}
    \mathrm{AoF} \label{eq:x3}.
\end{IEEEeqnarray}
\end{thm}
\begin{IEEEproof}
The proof is obtained via the straightforward application of Theorem 2 to the definitions of AoF and CQEI~\eqref{eq:x1}.
\end{IEEEproof}

% The derived expression for the arbitrary-order raw moments of the instantaneous signal-to-noise ratio completes the first-order statistical description of the model under consideration.
It is clear that AoF is an SNR-independent metric that is solemnly defined by the fading channel parameters; thus, CQEI is inversely proportional to the average SNR, which means that it linearly scales with SNR on a log-log plot.

Moreover, the AoF is a critical tool in the understanding of the fading channel regime, e.g., Rayleigh (if AoF equals to that of the Rayleigh channel, i.e., $\mathrm{AoF}=1$), hyper-Rayleigh, or lighter than Rayleigh (see, for instance, \cite{Gar19, Gvo23b}).

\subsubsection{Outage probability}\label{Sss-2-3-3}

The presented analytical work for the $\alpha$-BX-shadowed model performance is illustrated with the help of an outage probability as a quantifying metric. The $\mathrm{P_{out}}$ for a fixed predefined threshold $\bar{\gamma }_{th}$ (which guarantees the desired link quality or reliability) can be easily evaluated with the help of the obtained cumulative distribution function \eqref{eq:3}:
\begin{equation}\label{eq:7}
    \mathrm{P_{out}}=\mathbb{P}\{\gamma \leq\bar{\gamma }_{th}\}=F_{\gamma }(\bar{\gamma }_{th}).
\end{equation}

\begin{thm}
The outage probability of the wireless communications in case of the $\alpha$-BX-shadowed channel is given by
\begin{IEEEeqnarray}{rCl}\label{eq:8}
\mathrm{P_{out}} (\gamma_{th})&=&\frac{\left(\frac{m_Y\Omega_X}{m_Y\Omega_X+m_X\Omega_Y}\right)^{m_Y}\left(\frac{\gamma_{th}}{\bar{\gamma }}\right)^{\frac{\alpha m_X}{2}}}{\mathrm{C}_{\alpha }^{m_X}\Gamma (m_X+1)\exp\left(\frac{\gamma_{th}}{\bar{\gamma }}\right)^{\!\!\frac{\alpha }{2}}}  \times    \IEEEnonumber \\
&&\hspace{-60pt} \times    \Phi_2\!\!\left(\!1, m_Y; m_X; \frac{\left(\frac{\gamma_{th}}{\bar{\gamma }}\right)^{\!\!\frac{\alpha }{2}}}{\mathrm{C}_{\alpha }}\!,\left(\frac{\mathrm{C}_{\alpha }^{-1}m_X\Omega_Y \left(\frac{\gamma_{th} }{\bar{\gamma }}\right)^{\!\!\frac{\alpha }{2}}}{m_Y\Omega_X+m_X\Omega_Y}\right)\!\! \right),
\end{IEEEeqnarray}
\end{thm}
\begin{IEEEproof}
The proof can be obtained by combining outage probability definition \eqref{eq:7} with the presented first-order statistical description \eqref{eq:3} after some mathematical simplifications.
\end{IEEEproof}

Moreover, in practice, it is desirable to understand the asymptotic behavior of the system performance metric when the average SNR increases, i.e., for $\bar{\gamma }\to  \infty$. Thus, one can apply Theorem 1 to derive the upper and the lower bounds of the outage probability.

\begin{corollary}
The in the high-SNR region, the outage probability for the $\alpha$-BX-shadowed model can be upper and lower-bounded as follows
\begin{equation}\label{eq:9}
   \displaystyle \begin{cases}
     \displaystyle\mathrm{P_{out}^{LB}}=\mathrm{P_{out}^{UB}}e^{-\frac{1}{\mathrm{C}_{\alpha }}\left(\frac{\gamma }{\bar{\gamma }}\right)^{\frac{\alpha }{2}}}\leq\mathrm{P_{out}}\leq \mathrm{P_{out}^{UB}}, \\
   \displaystyle \mathrm{P_{out}^{UB}}=\frac{\left(\frac{m_Y\Omega_X}{m_Y\Omega_X+m_X\Omega_Y}\right)^{m_Y}}{\mathrm{C}_{\alpha }^{m_X}\Gamma (m_X+1)}\left(\frac{\gamma }{\bar{\gamma }}\right)^{\frac{\alpha m_X}{2}}.
  \end{cases}
\end{equation}
\end{corollary}
\begin{IEEEproof}
For such a scenario, one can notice that for a fixed $\gamma_{th}$, the increase of $\bar{\gamma }$ affects the exponent multiplier and the Appell function (i.e., $\Phi (   \cdot )\big|_{\bar{\gamma }\to  \infty}\to 1$). Thus, after some mathematical simplifications, the required result is obtained.
\end{IEEEproof}

% \begin{figure}[!t]
% \centerline{\includegraphics[width=\columnwidth]{fig1.eps}}
% \caption{Instantaneous SNR pdf for $\bar{\gamma }=0$~dB, $\Omega_X=\Omega_Y=2$~dB, $m_X=0.6$, $m_Y=0.5$, and various $\alpha$: solid lines - proposed analytical expression \eqref{eq:2}, markers - numeric simulation.}
% \label{fig1}
% \end{figure}

\begin{figure}[!t]
\centerline{\includegraphics[width=\columnwidth]{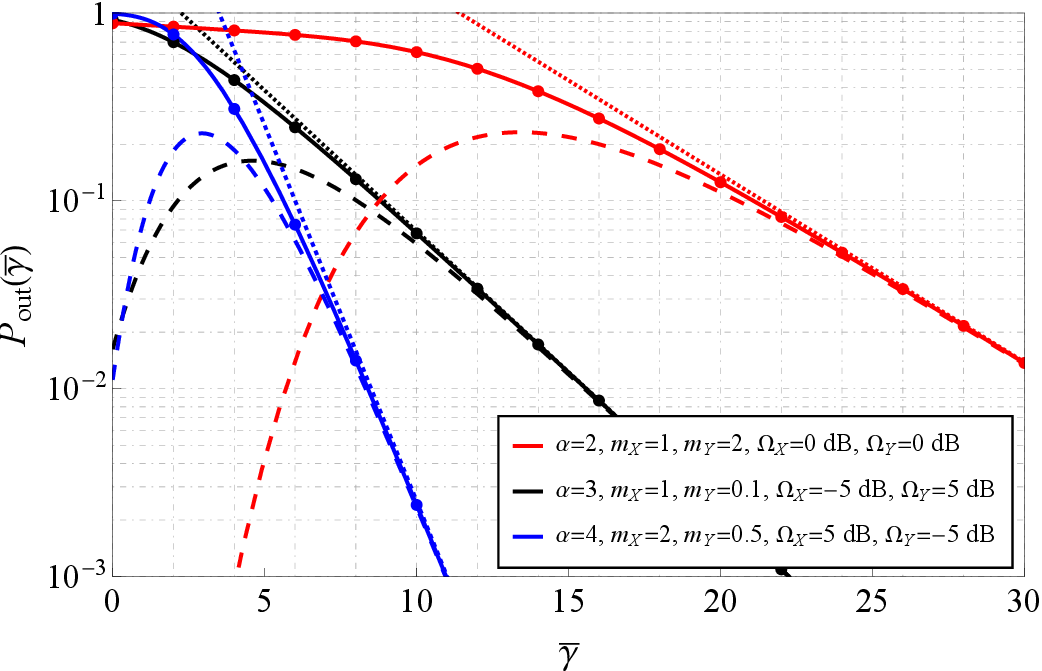}}
\caption{Outage probability for various fading scenarios: solid lines - proposed analytical expression \eqref{eq:8}, densely dashed - upper bound \eqref{eq:9}, loosely dashed - lower bound \eqref{eq:9}, markers - numeric simulation.}
\label{fig2}
\end{figure}

\begin{figure}[!t]
\centerline{\includegraphics[width=\columnwidth]{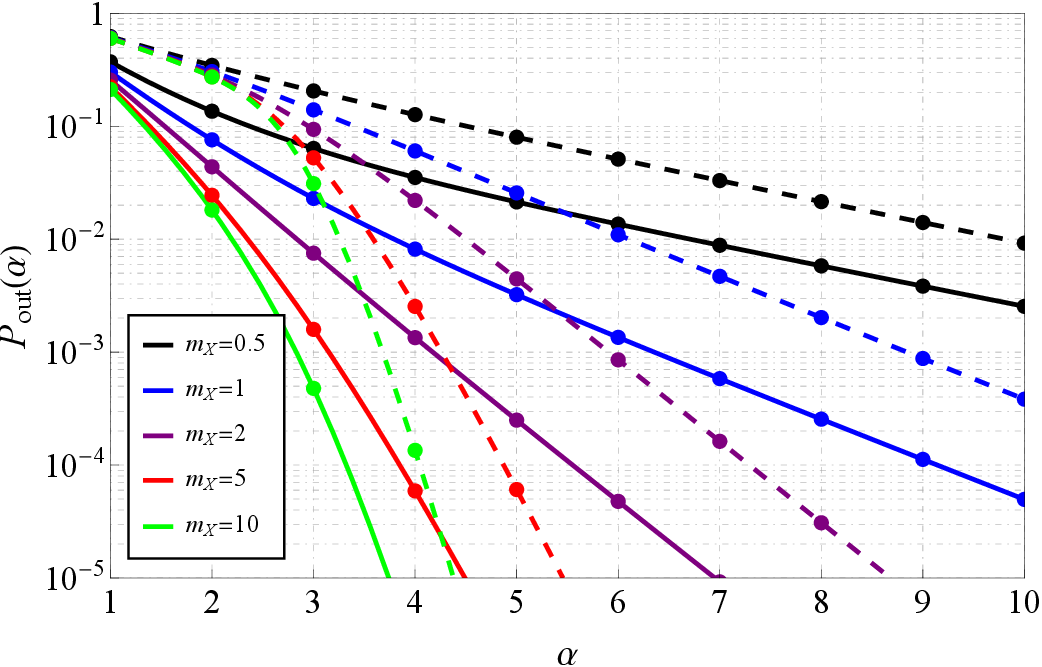}}
\caption {$\mathrm{P_{out}}$ versus $\alpha$ for $\gamma_{th}=3$~dB, $\bar{\gamma }=10$~dB, $\Omega_X=5$~dB, $\Omega_Y=-5$~dB: solid lines - $m_Y=2.5$, dashed lines - $m_Y=0.5$.}
\label{fig3}
\end{figure}

% \begin{figure}[!t]
% \centerline{\includegraphics[width=\columnwidth]{figx51.eps}}
% \caption {Amount of fading contour map
% for various $m_X, m_Y$ with $\alpha=2$, $\Omega_X=\Omega_Y=5$~dB, yellow dashed line - AoF for Rayleigh channel.}
% \label{fig51}
% \end{figure}

\subsubsection{Joint quality-reliability analysis}\label{Sss-2-3-4}
It is well known that there is no single metric that can solemnly characterize the fading communication channel. Thus, it is proposed to assess the system performance in terms of the joint quality-reliability analysis. For later on, one assumes outage probability $\mathrm{P_{out}}$ (derived herein) as the primary metrics quantifying link reliability, and the average bit error rate $\mathrm{\bar{P}_{err}}$ (derived in closed form in \cite{Gvo23}) – quantifying its quality. So the deviation of a joint $\mathrm{P_{out}}-\mathrm{\bar{P}_{err}}$ curve from a bisecting line for all possible $\bar{\gamma }$ (or $\mathrm{CQEI}$) demonstrates the preference of quality over reliability (or vice versa) for a given fading conditions. Thus, such an analysis can provide some insights into the possible parameters' optimization.

\section{Simulation and results}\label{S-3}
\begin{figure}[!t]
\centerline{\includegraphics[width=\columnwidth]{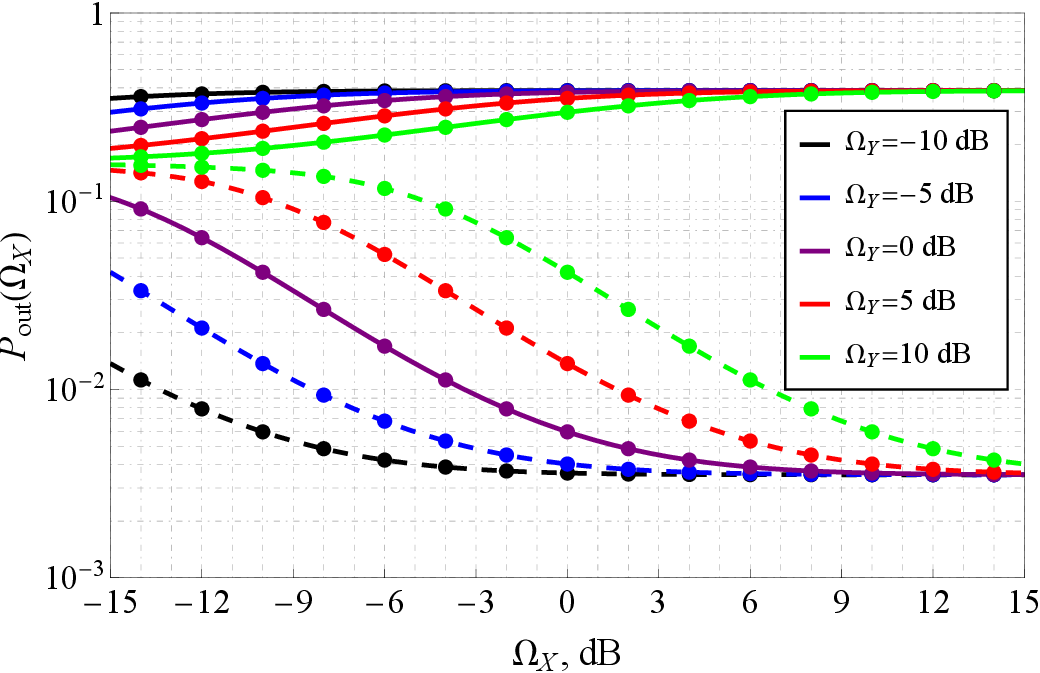}}
\caption {$\mathrm{P_{out}}$ versus $\Omega_X$ for $\gamma_{th}=3$~dB, $\bar{\gamma }=10$~dB, $\alpha=3.5$: solid lines - $m_X=0.2, m_Y=0.5$, dashed lines - $m_X=2.2, m_Y=0.5$.}
\label{fig4}
\end{figure}

\begin{figure}[!t]
\centerline{\includegraphics[width=\columnwidth]{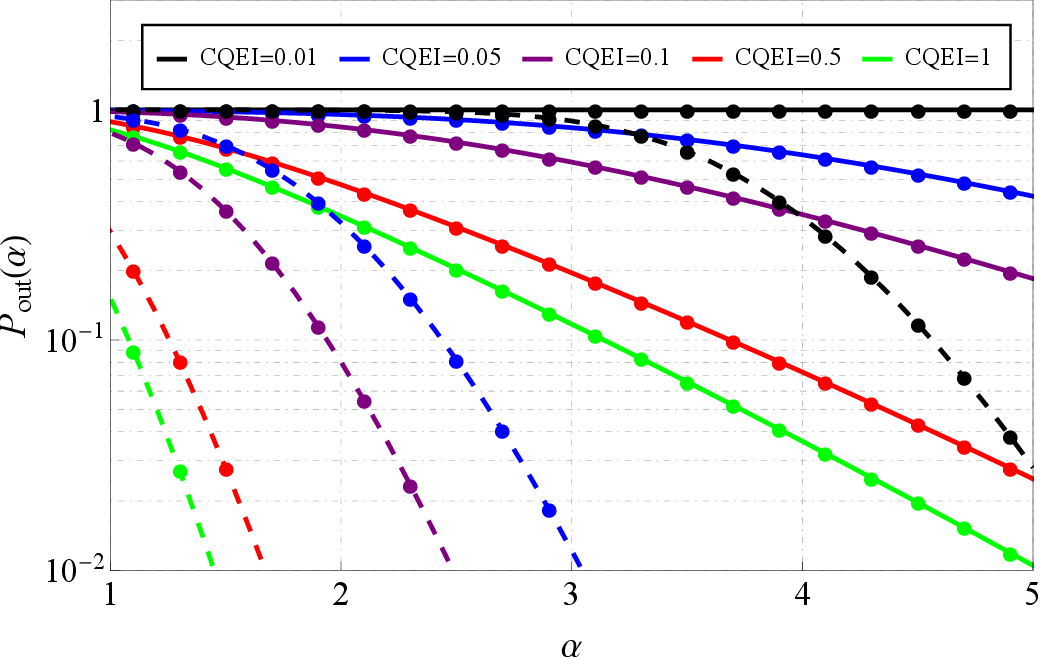}}
\caption {$\mathrm{P_{out}}$ versus CQEI for $\gamma_{th}=10$~dB: solid lines - $\Omega_X=\Omega_Y=0$~dB and $m_X=m_Y=0.5$, dashed lines - $\Omega_X=\Omega_Y=5$~dB and $m_X=m_Y=3$.}
\label{fig5}
\end{figure}

\begin{figure}[!t]
\centerline{\includegraphics[width=\columnwidth]{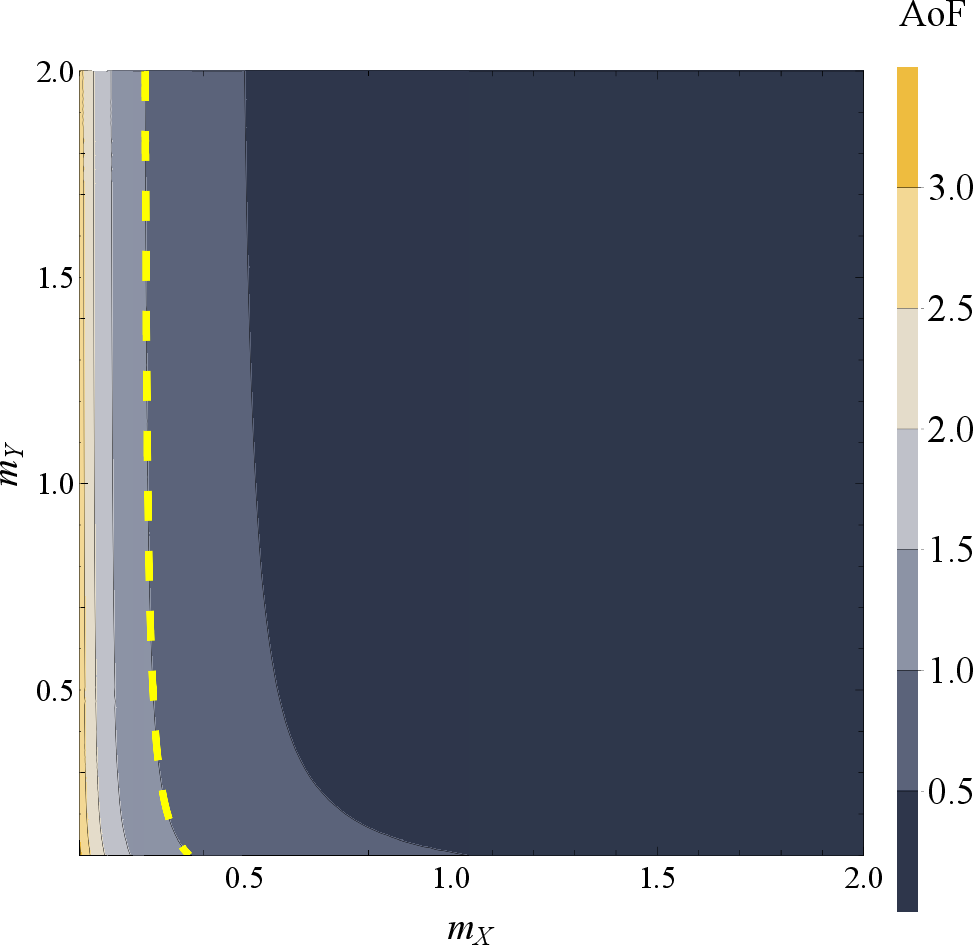}}
\caption {Amount of fading contour map
for various $m_X, m_Y$ with $\alpha=4$, $\Omega_X=\Omega_Y=5$~dB, yellow dashed line - AoF for Rayleigh channel.}
\label{fig52}
\end{figure}

\begin{figure}[!t]
\centerline{\includegraphics[width=\columnwidth]{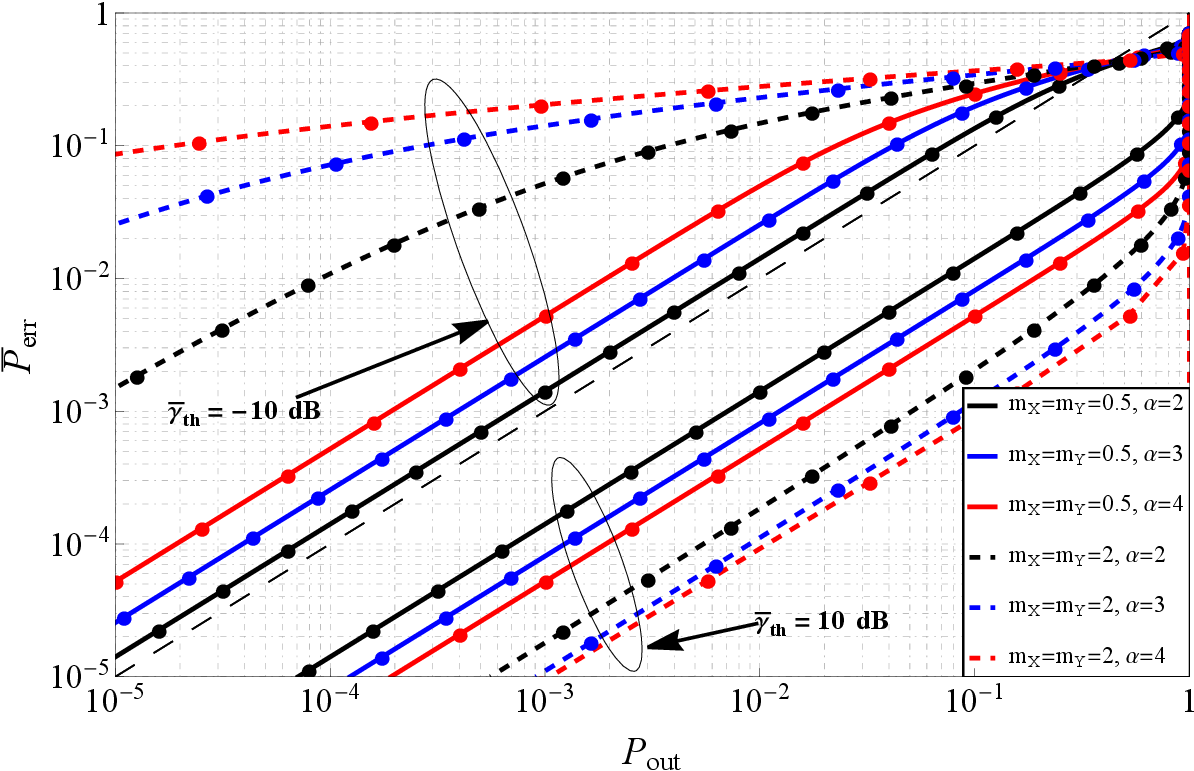}}
\caption {Joint quality-reliability plot for the case of $\Omega_X=\Omega_Y=-5$~dB, QAM-16 modulation, and two $\gamma_{th}$ (i.e., $\gamma_{th}=-10$~dB and $\gamma_{th}=10$~dB); black dashed line without markers denotes the parity between link quality and reliability.}
\label{fig6}
\end{figure}

To assess the outage performance of the $\alpha$-BX-shadowed channel for various parameters, numerical analysis was deployed. The consistency of the derived expressions was tested with the help of the Monte-Carlo simulation, which was performed as follows: on the first stage, $10^7$ samples of the signal envelope corresponding to the initial BX model were obtained (see expressions (2)-(4), in \cite{Sil21}), which were further modified to account for the LoS  fluctuations (see \cite{Olu20a}), transformed to assume nonlinear distortions (see \cite{Yac07b}), and recalculated into the instantaneous SNR samples.  In \cite{Gvo23}, it was demonstrated that such a simulation procedure yields results that are in full compliance with the probabilistic model and are in good agreement with the performed herein numerical analysis (see Fig.~\ref{fig2}-\ref{fig6}).

To analyze the outage probability \eqref{eq:8} (evaluated in terms of the derived cdf \eqref{eq:3}), all the possible channel conditions were considered, e.g., heavy/weak dominant/multipath components with strong/light fading (see Fig.~\ref{fig2}-\ref{fig4}). Moreover, the closed-form results were accompanied by the derived upper/lower bounds \eqref{eq:9} (see Fig.~\ref{fig2}). It can be seen that the average SNR, starting from which \eqref{eq:9} delivers a reasonably good approximation, depends on $\alpha$: the greater the parameter value the smaller this SNR is (e.g., $20$~dB for $\alpha=2$, $10$~dB for $\alpha=3$, and  $7$~dB for $\alpha=4$).  Fig.~\ref{fig3} and \ref{fig4} demonstrate the relation of $\mathrm{P_{out}}$ with the parameters of the model. It was observed that the greatest impact (besides $\alpha$) is induced by the overall fading parameter ($m_X$). Moreover, depending on its value, the  system performance may improve or degrade irrespective of other parameters. For instance (see Fig.~\ref{fig4}), the increase of $\Omega_X$ induces the growth of $\mathrm{P_{out}}$ in case of small $m_X$ and the decrease of $\mathrm{P_{out}}$ in case of large $m_X$.

The derived expression for the AoF (see Theorem 3) helps connect $\mathrm{P_{out}}$ with $\mathrm{CQEI}$ and execute  the outage analysis in terms of the CQEI (see Fig.~\ref{fig5}). One can note that the impact of $\alpha$ is more pronounced for channels with better propagation conditions (i.e., $\Omega_X=\Omega_Y=5$~dB and $m_X=m_Y=3$ in Fig.~\ref{fig5}, contrary to  $\Omega_X=\Omega_Y=0$~dB and $m_X=m_Y=0.5$). Moreover, the impact of $\mathrm{CQEI}$ is nonequal for the fixed relative change of $\mathrm{CQEI}$, e.g., the decrease of $\mathrm{CQEI}$ by the order of magnitude from $1$ to $10^{-1}$  can be compensated by reducing $\alpha$ from $2.5$ to $1.4$ (keeping $\mathrm{P_{out}}$ fixed on a level of $10^{-2}$), whereas the same decrease of $\mathrm{CQEI}$ from $10^{-1}$ to  $10^{-2}$ requires a much greater change in $\alpha$.

As it was mentioned earlier, the analysis of the AoF itself helps to identify the regime of the propagation channel. The yellow line in Fig.~\ref{fig52} corresponds to the Rayleigh fading, detaching the hyper-Rayleigh from lighter-than-Rayleigh fading and defining the set of fading/shadowing parameters (i.e., $m_X, m_Y$) that constitute those scenarios. It must be pointed out that hyper-Rayleigh fading is quantified by three parameters (i.e., AoF, $\mathrm{P_{out}}$, and channel capacity). Since the performed herein analysis operates with the AoF only, one can define the cases of no more than weak hyper-Rayleigh (see the area to the left from the yellow line in Fig.~\ref{fig52}). The obtained results demonstrate the uneven contribution of $m_X$ and $m_Y$ (in favor of $m_X$) to the specific regime definition, e.g., for $\alpha=4$, the weak hyper-Rayleigh fading is observed in case of $m_X<0.25$ for any values of $m_Y$. The set of parameters $(m_x, m_Y)$, which correspond to the hyper-Rayleigh scenario, heavily depend on the nonlinear transformation coefficient $\alpha$.

Lastly, the proposed joint quality-reliability analysis was performed for the assumed channel model (see Fig.~\ref{fig5}). For the case study, a communication system with QAM-16 modulation was employed. The average bit error rate was evaluated with the help of the expression $(10)$ from \cite{Gvo23} truncated to $4$ terms. In Fig.~\ref{fig5}, the dashed black bisector (without markers) represents the state of parity (defined as a set of the channel parameters) between $\mathrm{P_{out}}$ and $\mathrm{\bar{P}_{err}}$. It can be seen that the main factor that influences the balance between the two is the threshold average SNR (defined for $\mathrm{P_{out}}$). Moreover, the increase of $\alpha$ further enhances this imbalance. It was also identified that for a very poor channel (i.e., low quality and reliability), the balance is biased in favor of  $\mathrm{P_{out}}$.

\section{Conclusion}\label{S-4}

The research performs the outage analysis of the $\alpha$-BX-shadowed model. For the given channel, the main statistical functions (i.e., cdf, pdf,  and raw moments) of the instantaneous SNR are derived.  The exact expressions of the amount of fading, channel quality estimation indicator, and the outage probability are evaluated, and its lower and upper bounds are estimated. The numerical analysis conducted confirmed the accuracy of the analytical work and aided in examining the dependence of the outage probability on all the possible channel parameters.

\ifCLASSOPTIONcaptionsoff
  \newpage
\fi

\end{document}